\documentclass[12pt]{article}
\usepackage{latexsym}
\usepackage{amssymb}
\usepackage[english]{babel}
\usepackage{amsfonts}
\usepackage{graphicx}
\usepackage{amscd}

\newtheorem{theorem}{Theorem}

\newtheorem{definition}{Definition}

\newtheorem{remark}{Remark}

\newcommand{\A}{\mathcal{A}}

\newcommand{\amm}{a^{(r)\mu\dag}_{m}}
\newcommand{\ann}{a^{(s)\nu\dag}_{n}}

\newcommand{\Bc}{\mathfrak{B}_{c}}

\newcommand{\Bo}{\mathfrak{B}}

\newcommand{\C}{\mathbb{C}}

\newcommand{\HH}{\mathcal{H}}
\newcommand{\Ma}{\mathfrak{M}}
\newcommand{\N}{\mathbb{N}}
\newcommand{\nn}{\mathcal{N}}

\newcommand{\QQ}{\mathcal{Q}}
\newcommand{\RR}{\mathbb{R}}
\newcommand{\Sc}{\mathcal{S}}

\newcommand{\K}{K_{0}}

\newcommand{\Z}{\mathbb{Z}}

\newcommand{\Cs}{C^{*}}

\newcommand{\tP}{\widetilde{\Phi}}

\begin{document}
\begin{flushright}
SISSA 14/2004/fm
\end{flushright}
\begin{center}
{\huge\textbf{K-theory in cutoff version of Vacuum String Field Theory      \\[1cm]}}
\small{A. Parodi \footnote{adryparodi@libero.it; aparodi@sissa.it}\\[0.5cm]}
\small{\emph{I.S.A.S. International School for Advanced Studies, \\
Via Beirut 2-4, Trieste, Italy}\\[3cm]}
\end{center}
{\center\large\textbf{Abstract\\[1cm]}}
Solutions of the Vacuum String Field Theory (VSFT) equation of motion involving matter part are given by projectors, and they represent nonperturbative solutions (e.g. the sliver) interpreted as $D$25-branes (or lower dimensional branes), but they are not mathematically well defined as they have zero norm. In this work we will use a regularization procedure based on the cutoff version of Moyal String Field Theory (MSFT), a particular version of VSFT, and we will see that both the sliver and the butterfly states, in this regime, have a good mathematical description. In particular they are exponential functions
belonging to $\Sc(\RR^{2Nd})$, the space of Schwartzian functions equipped with the $*$-product. Then we  prove that if we classify those regularized solutions with K-theory group built out of the $\Cs$-algebra $\overline{\Sc}(\RR^{2Nd})$ we find exactly the same result obtained considering a K-theoretic classification of $D$25-branes in usual string theory, using the topological K-theory of vector bundles over the $D$25-brane worldvolume. We then comment on the meaning of this result and possible physical implications.
\newpage
\tableofcontents
\section{Introduction}
\label{sec:Introduction}
$D$-branes are fundamental objects needed to understand the non perturbative aspects of string theories. They are especially used to test various duality relations between different string theories, so their analysis deserves a very particular attention and care.\\
We can study them in many different ways, both in closed string theories and open string theories, and depending on what specific framework we consider, different properties of these objectis arise.\\
In a first quantized string theory they can be introduced only as bounduary conditions, but if we use a second quantized string theory, more precisely Closed String Field Theory, we expect that they arise as particular solitonic solutions and this fact was analyzed quite recently in \cite{[54],[55]}.\\ 
In this work we want to focus on $Dp$-branes using open string theory. This can be done using Sen's conjecture about the decay of unstable $Dp$-branes into the closed string vacuum or lower dimensional branes. In particular if we consider Open Bosonic String Field Theory (OSFT) on a $Dp$-brane we can obtain lower dimensional $Dq$-branes as solutions of the equations of motion of tachyonic effective theory obtained from OSFT.\\ 
OSFT is the best known string field theory (described for the first time by Witten in \cite{[20]}), but due to the great complexity of equations of motions, so far no analytic solutions are known, only numerical ones. This means that solutions representing lower dimensional branes are given by numerical expressions (see \cite{[18]},\cite{[13]} and references therein).\\
Fortunately a more manageable description of OSFT was found in \cite{[14],[19]}, where the authors performed a (singular) string field redefinition and built string field theory around the closed string vacuum. In this case the kinetic operator is made of pure ghost and the equations of motion are much easier: using the simplifying ansatz 
\[\Psi=\Psi_{m}\otimes \Psi_{g}\]
where $\Psi_{m}$ is a matter string field and $\Psi_{g}$ is a ghost string field, equations of motion for the matter part become 
\[\Psi_{m}* \Psi_{m}=\Psi_{m}\]
that is to say, the equation for a projector. Some analytic solutions were found, for example the sliver $\Xi$ representing a $D$25- brane so as butterfly states described for the first time in \cite{[44]}, analyzed in some details in \cite{[45],[46]} from CFT point of view and \cite{[32]} from oscillator point of view. So we have that VSFT admits solutions representing classical solitonic objects as the theory itself, at this level, is a classical theory, not including quantum corrections. Other solutions representing lower dimensional $Dp$-branes were found in \cite{[14]}\\
\\
It is well known that $D$-brane configurations can be classified in usual string theory by means of K-homology and K-cohomology, both based on vector bundles carried by the $D$-branes configuration. While in superstring theories this classification is very interesting from a physical point of view as it determines the Ramond-Ramond charges of the system, in bosonic theory it seem to have only a mathematical meaning, as bosonic $D$-branes do not carry any conserved charge and so they are expected to be unstable. In particular we focus on the classification based on  topological K-theory groups (K-cohomology). In our work we will consider the simplified framework of flat space-time and flat branes.\\
We would like to see if is it possible to obtain the same informations on $Dp$-branes seen as objects arising directly in string theory, that is, as solutions of some String Field Theory. This can be achieved in VSFT.\\
As VSFT admits multiple $Dp$-brane solutions, obtained adding orthogonal projectors (see \cite{[15]}), it is an interesting task to try to classify such configurations by means of algebraic K-theory groups (which are obtained using orthogonal projectors), comparing such classification with the topological one discussed above, and checking if we gain some new physical information in this framework.\\
In this work we will consider only $D$25-brane solutions represented explicitly by the sliver and the butterfly. Unfortunately these solutions are singular as they have zero norm, so they need to be regularized. A particularly useful regularization is obtained in Moyal String Field Theory (this version of VSFT describes Witten's product as a Moyal product over an infinite dimensional manifold with coordinates $\left\{x_{2n},p_{2n}\right\}_{n\geq 1}$, so we have a well defined mathematical structure, see \cite{[35]}), where it is employed considering a cutoff, that is, only a finite number $2N$ of coordinates $\left\{x_{2},\ldots,x_{2N},p_{2},\ldots,p_{2N}\right\}$. In this language both the sliver and the butterfly have a finite norm and they are represented by Gaussian functions. This is a very interesting fact as Gaussian functions belong to the $\Cs$-algebra given by the completion of Schwartzian functions $\Sc(\RR^{2Nd})$ equipped with the Moyal product. Given a $\Cs$-algebra it is a well known procedure to assign an algebraic K-theory group to it, considering equivalence classes of projectors. As both the butterfly and the sliver cutoff versions are projectors we can classify them using K-theory. Of course we cannot have in principle a complete information on the full solutions as we are considering a cutoff enviroment, but nevertheless it can be a good starting point for future developements aiming to solve the full problem.\\
The main part of this work consists in determining this classifying algebraic K-theory group and comparing it with the topological one determined in usual string theory.\\
\\
Of course this wants to be a preliminar analysis carried over in the simplified framework of a bosonic string field theory. The real interesting test is to apply all this machinery to superstring theories, where K-theory has a well defined physical meaning in the classification of $Dp$-branes (see \cite{[11],[12]}). In particular it is interesting to check if, starting from Superstring Field Theory (eventually Vacuum Superstring Field Theory) solutions representing $Dp$-branes, we manage to classify them by the same K-theory group as in \cite{[11]}. This is beyond the scope of this work.\\
\\
As the quest for all projectors of a $\Cs$-algebra can be very complicated, it can be useful to check if the $\Cs$-algebra we are using is isomorphic to some $\Cs$-algebra whose K-theory group is known.
In particular we use the fact that the $\Cs$-algebra $\overline{\Sc}(\RR^{2Nd})$, obtained from the norm completion of $\Sc(\RR^{2Nd})$, is isomorphic to the $\Cs$-algebra of compact operators over a separable Hilbert space. This result is very important for us as the K-theory group of compact operators is well known.\\
\\
The paper is organized as follows: in section 2 we recall briefly the Vacuum String Field Theory construction and in particular the form of the sliver solution representing a $D25$-brane, focusing on some mathematical problems that plague it, and that require a cut off in order to be solved. In section 3 we will analyse MSFT in some details, comparing it with the oscillator construction and focusing on the cutoff, especially for the sliver and the butterfly. In section 4 we review some standard material on the relation between Schwartzian functions and compact operators, in particular the result we need is  $\overline{\Sc}(\RR^{2Nd})\simeq \Bc(L^{2}(\RR^{Nd}))$. In section 5 we recall very briefly the construction of K-theory groups starting from projectors of a $\Cs$-algebra. Finally in the last section we comment on results found, especially on the role of K-theory in the description of bosonic $D$-25-brane. We will also highlight possible future directions.
\section{$D$-branes in VSFT}
\label{sec:DBranesInVSFT}
Let us briefly remind that Open String Field Theory (OSFT) is described by the action 
\begin{equation}\label{eq46.ps}
	S=-\frac{1}{g^{2}_{0}}\int\Big(\frac{1}{2}\Phi*_{W}Q_{B}\Phi+\frac{1}{3}\Phi*_{W}\Phi*_{W}\Phi\Big)
\end{equation}
where $g_{0}$ is a dimesionful parameter and $Q_{B}$ is the usual BRST charge arising in first quantization (in particular it mixes matter and ghost part). Here $*_{W}$ is the so called \emph{Witten product}, and it is an associative, non commutative product describing the gluing of two strings (see \cite{[20]}).\\
Equations of motion are then given by
\begin{equation}\label{eq300}
	Q_{B}\Phi+\Phi*_{W}\Phi=0
\end{equation}
Here $\Phi\in \A$ is a generic string field, where $\A$ is the algebra of string fields. We want to emphasize that the original description of OSFT was quite abstract, all the operations on the algebra $\A$ had a precise physical meaning (like gluing and interaction of strings) but they had to be translated in well defined mathematical expressions.\\
A possible concrete approach was given in \cite{[36]} where both $Q_{B}$ and $*_{W}$ were written down in terms of Fock oscillators.\\
It would be interesting to find analytic solutions of this equation of motion, but due to the mixing of matter and ghost part encountered in the kinetic term described by the BRST charge it is very difficult to solve it, and so far only numerical solutions are aviable. In particular by means of a truncation level procedure in OSFT (see \cite{[13]}) we manage to find some numerical solutions that, according to Sen's conjecture are interpreted as lower dimensional $Dp$-branes.\\
\\
In order to find analytic solutions in string field theory framework we can change point of view.\\
In the work \cite{[19]} it was shown that we can shift OSFT around a particular vacuum, called Closed String Vacuum, and redefine string fields centered around it, obtaining another version of the theory, called Vacuum String Field Theory (VSFT). Its equation of motion is much easier to solve than original OSFT, as shown in \cite{[14]} and related works, as we choose a new BRST operator $\QQ$ made only of ghosts. In particular this theory admits $Dp$-branes solutions which can be described in a closed form.\\
\\
In this section we will decribe briefly how VSFT is built and how $D25$-brane solutions are constructed, following the seminal works \cite{[14],[19]}. In particular we use the Fock space language although other descriptions like half string (see \cite{[15],[31]}) and CFT (see \cite{[48]}) can be used. This construction can also be used  to determine solutions representing lower dimensional $Dp$-branes, but as we are interested only in $D25$-branes we will not describe them.
\subsection{General setup}
\label{sec:GeneralSetup}
Using Sen's conjecture about tachyon condensation we are able to find a numerical solution $T=\Phi_{0}$ of OSFT representing the closed string vacuum and whose components are the vacuum expectation values of many scalar fields. Infact OSFT is constructed on the open string vacuum corresponding to the string field $\Phi=0$. The main idea of \cite{[14],[19]} is to build a new string field theory, centered around the closed string vacuum, so the authors expand a generic string field $\Phi$ around $\Phi_{0}$ as 
\begin{equation}\label{eq45.ps}
\Phi=\Phi_{0}+\widetilde{\Phi} 
\end{equation}
and then considering the fluctuations $\widetilde{\Phi}$ around closed string vacuum, using the appropriate kinetic term. Unfortunately we are not able to analyze OSFT around $\Phi_{0}$ as we do not know the analytic form of this field.\\
A useful idea has been carried over in \cite{[19]} where the authors expanded the action around $\Phi_{0}$, plugging (\ref{eq45.ps}) in the usual OSFT action (\ref{eq46.ps}) and finding
\begin{equation}
	S(\Phi_{0}+\widetilde{\Phi})=S(\Phi_{0})-\frac{1}{g^{2}_{0}}\Big[\frac{1}{2}\langle \tP, Q\tP\rangle+\frac{1}{3}\langle\tP,\tP *_{W}\tP\rangle\Big]
\end{equation}
with
\begin{equation}
	Q\tP=Q_{B}\tP+\Phi_{0}*_{W}\tP+\tP*_{W}\Phi_{0}
\end{equation}
Now we consider a field redefinition that simplifies drammatically our action:
\begin{equation}
	\tP=e^{K}\Psi
\end{equation}
where $K$ is an appropriate operator, and so we obtain a new action for fields $\Psi$:
\begin{equation}
	S(\Psi)=-\frac{1}{g^{2}_{0}}\Big[\frac{1}{2}\langle \Psi, \QQ \Psi\rangle+\frac{1}{3}\langle\Psi,\Psi *_{W}\Psi\rangle\Big]
\end{equation}
with the new kinetic term 
\begin{equation}
	\QQ=e^{K}Q e^{-K}
\end{equation}
In particular it was postulated that the operator $\QQ$ has a very simple form, made only of ghosts operators in order to have trivial cohomology. This means that this theory does not admit perturbative open string solutions, described in first quantization by BRST cohomology. So we expect to find only non perturbative solutions like $Dp$-branes around the closed string vacuum.\\
This new description of SFT is called Vacuum String Field Theory (VSFT).\\
Equations of motion of this new action are given by
\begin{equation}
	\QQ \Psi+\Psi*_{W}\Psi=0
\end{equation}
It was argued in \cite{[14]} that fields $\Psi$ can be factorized in a matter and a ghost part: 
\[\Psi=\Psi_{m}\otimes \Psi_{g},\] 
so that, using the fact that $\QQ$ is made of pure ghost operators, we have factorized equations of motion
\begin{equation}
	\QQ \Psi_{g}=-\Psi_{g}*^{g}\Psi_{g}
\end{equation}
and
\begin{equation}\label{eq305}
	\Psi_{m}*^{m}\Psi_{m}=\Psi_{m}
\end{equation}
where $*^{g},*^{m}$ denote the multplication rules in the ghost and matter sector respectively.\\
\\
It was pointed out in \cite{[14]} that the ghost part $\Psi_{g}$ can be chosen to be the same for all solutions, so let us deal with it first. The work \cite{[49]} tells us that if we choose string fields belonging to the Siegel gauge,
the kinetic term becomes
\begin{equation}
	\QQ=c_{0}+\sum_{n>0}(-1)^{n}(c_{2n}+c_{-2n})
\end{equation}
while the ghost equation of motion is
\begin{equation}
	\Psi_{g}*^{g}\Psi_{g}=\Psi_{g}
\end{equation}
A squeezed state solution of this equation is given by
\begin{equation}\label{eq120.ps}
	\vert \Psi_{g}\rangle=\left\{\det(1-\tilde{X})^{\frac{1}{2}}\det(1+\tilde{T})^{\frac{1}{2}}\right\}\exp\Big(\sum_{n,m=1}^{\infty}c^{\dag}_{n}\tilde{S}_{nm}b^{\dag}_{m}\Big)c_{1}\vert \tilde{0}\rangle
\end{equation}
where the matrix $\tilde{S}$ is given by the expression $\tilde{S}=C\tilde{T}$, with
\begin{equation}
	\tilde{T}=\frac{1}{2\tilde{X}}\Big(1+\tilde{X}-\sqrt{(1+3\tilde{X})(1-\tilde{X})}\Big)
\end{equation}
Here $C=(-1)^{n}\delta_{nm}$, $\tilde{X}=C\tilde{V}^{11}$, and $\tilde{V}^{11}$ is one of the Neumann matrices that appear in the ghost tree point vertex $\vert V_{3}\rangle$ describing Witten's product $*^{g}$ in Fock space language.\\
This solution is taken as universal, that is, once we have any matter solution $\vert \Psi_{m}\rangle$ satisfying matter projector equations of motion (\ref{eq305}), the full solution is given by $\vert \Psi\rangle=\vert \Psi_{m}\rangle\otimes\vert \Psi_{g}\rangle$ where $\vert \Psi_{g}\rangle$ has the above form.\\
\\
Now we focus on the matter part. It is clear from (\ref{eq305}) that in VSFT, using the factorization ansatz, solutions of the equations of motion for the matter part are given by projectors.\\
The first attempt to find an analytic solution to string field theory was made in \cite{[26]}, where it was assumed that the solution is a squeezed state of the form
\begin{equation}
	\vert \Psi\rangle=\mathcal{N}^{26}\exp\Big(-\frac{1}{2}\sum^{\infty}_{m,n=1}a^{\dag}_{m}S_{mn}a^{\dag}_{n}\Big)\vert 0\rangle
\end{equation}
where $\mathcal{N}$ is a normalization factor and $S_{mn}$ is a symmetric, infinite dimensional matrix. Both these quantities must be determined solving equations of motion, that in our case are equations for a projector in the same way as above for the ghost part.
\subsection{The Sliver solution}
\label{sec:TheSliverSolution}
OSFT can be considered over a generic background, but in order to simplify calculations we consider OSFT over a flat $D$25-brane with a Euclidean metric. This is needed in order to have well behaved Gaussian functions in the following. Moreover in flat spacetime we can use Fock space language.\\
Witten's product in the matter part assumes the following general form:
\begin{equation}\label{eq169.ps}
	\vert V_{3}\rangle=\int d^{26}p_{(1)}d^{26}p_{(2)}d^{26}p_{(3)}\delta^{(26)}(p_{(1)}+p_{(2)}+p_{(3)})\exp(-E)\vert 0,p\rangle_{123}
\end{equation}
where
\begin{equation}
	E=\frac{1}{2}\sum_{{r,s \atop n,m\geq1}}\delta_{\mu\nu}\amm V_{mn}^{rs}\ann+\sum_{{r,s \atop m\geq1}}\delta_{\mu\nu}p^{\mu}_{(r)}V^{rs}_{0n}\ann+\frac{1}{2}\sum_{r}\delta_{\mu\nu}p^{\mu}_{(r)}V^{rr}_{00}p^{\nu}_{(r)}
\end{equation}
Here $r,s=1,2,3$ specify the string in the three strings vertex.\\
Now we look for a $p$-independent solution of matter equation of motion, so we consider only 
\begin{equation}\label{eq123.ps}
	E=\frac{1}{2}\sum_{{r,s \atop n,m\geq1}}\delta_{\mu\nu}\amm V_{mn}^{rs}\ann
\end{equation}
It was found in \cite{[14]}, that there exists a space-time independent solution $\vert\Xi_{m}\rangle$ called \emph{sliver} that can be written in closed form, which is interpreted as a $D$25-brane. The algebraic expression in terms of Fock state oscillators is given by
\begin{equation}
	\vert \Xi_{m}\rangle=\left\{\det(1-X)^{\frac{1}{2}}\det(1+T)^{\frac{1}{2}}\right\}^{26}\exp\Big(-\frac{1}{2}\delta_{\mu\nu}\sum_{m,n\geq1}S_{mn}a^{\mu\dag}_{m}a^{\nu\dag}_{n}\Big)\vert 0\rangle
\end{equation}
where 
\[S=CT~~~~;~~~~~~T=\frac{1}{2X}\Big(1+X-\sqrt{(1+3X)(1-X)}\Big)\]
and $C_{mn}=(-1)^{m}\delta_{mn},~m,n\geq1$, $X=CV^{11}$.\\
So the global solution, including the universal ghost part $\vert\Psi_{g}\rangle$ is given by
\begin{equation}\label{eq125.ps}
\vert\Xi\rangle=\vert\Psi_{g}\rangle\otimes \vert\Xi_{m}\rangle.
\end{equation}
Let us now focus on the sliver solution, to uderstand why we need to introduce a cutoff. As shown above it is written in terms of oscillators acting on the Fock vacuum, but it can be shown it does not belong to the Fock space, as it has vanishing norm:
\begin{equation}
	\langle \Xi_{m}\vert \Xi_{m}\rangle=\frac{V^{(26)}}{(2\pi)^{26}}(\det(1-X)^{3/4}\det(1+3X)^{1/4})^{26}
\end{equation}
where we used
\[\langle 0\vert 0\rangle=\delta^{(26)}(0)=\frac{V^{26}}{(2\pi)^{26}}\]
So, up to an infinite factor arising from the infinite 26-dimensional volume we find the product of two determinants. Both from numerical and analytic considerations (see \cite{[37]}) we have that the above product of determinants is zero. Such result is found  regularizing the theory, considering only a finite number $L$ of oscillators (in this case we get a finite, non vanishing answer), and then taking the limit $L\rightarrow \infty$. In a first time it was thought that the inclusion of the universal ghost part $\vert \Psi_{g}\rangle$ manages to give a finite, nonvanishing answer, but from the regularization used in \cite{[37]} it seems that it is not true: the full sliver state $\vert \Xi\rangle=\vert \Xi_{m}\rangle\otimes\vert \Psi_{g}\rangle$ still has vanishing norm, so it does not belong to the Fock space of open string.\\
We recall that the sliver was the first solution found in VSFT, but nowadays other solutions are known, in particular the butterfly. These solutions too are expressed in terms of squeezed states if we use the Fock space language. For an introduction to the butterfly state see \cite{[44]},\cite{[45]},\cite{[46]}. It has the same factorized form (\ref{eq125.ps}), where the ghost part has the universal form (\ref{eq120.ps}) and it is plagued by the same problem of the vanishing norm as the sliver.\\
This is a big trouble if we want to employ construction outlined in \cite{[1]}, that tries to see VSFT solutions as elements of a $\Cs$-algebra, infact elements in a $\Cs$-algebra have nonvanishing, finite norm. \\
Our task is to try to classify VSFT solutions by means of K-theory groups and, as suggested in \cite{[1]} we can obtain them starting from a $\Cs$-algebra containing such solutions. So it seems a good starting point to consider a cutoff version of VSFT which was built in \cite{[35]}: from above considerations it seems that the problems arise when we take an infinite number of oscillators (that is the $L\rightarrow \infty$ limit), so we can expect to have a mathematically well posed problem, that is to say, states with a finite non vanishing norm, once we consider a finite number of degrees of freedom.
\section{Bacics of MSFT}
\label{sec:BscicsOfMSFT}
In this section we want to describe the basic ideas underlying the description of VSFT by means of Moyal product, which is a well known mathematical structure. We recall from \cite{[35]} all the ingredients we need in order to have a self-contained discussion. A much more detailed analysis of the construction can be found in \cite{[33],[34],[35]}.
\subsection{Moyal's product in VSFT}
\label{sec:MoyalSProductInVSFT}
If we consider a generic string configuration
\begin{equation}
	x^{\mu}(\sigma)=x^{\mu}_{0}+\sqrt{2}\sum_{n=1}^{\infty}x^{\mu}_{n}\cos(n\sigma)
\end{equation}
the degrees of freedom are the $\left\{x_{0},x_{n}\right\}$ coefficients, and they are an infinite number. A string field is described in position representation as $\psi(x_{0},x_{2n},x_{2n-1})=\langle x\vert \psi\rangle$, where $\vert \psi\rangle$ is the string field in Fock space formalism. The major part of the theory does not need that we specify the number of oscillators, so we can consider a finite number $2N$ of them, employing the cutoff, as we will see in details later on. Usual VSFT is obtained considering $2N\rightarrow \infty$. Only some special structures need the full infinite number limit, but we will not use them.\\
Now we can formulate VSFT in Moyal language performing a Fourier transform on the odd position modes $\left\{x_{2n-1}\right\}$ of the string:
\begin{equation}
	\Psi(\bar{x},x_{2n},p_{2n})=\det(2T)^{d/2}\int\prod_{n=1} dx_{2n-1} e^{{\frac{2i}{\vartheta}\sum_{n,m}}p_{2n}T_{2n,2m-1}x_{2m-1}}\psi(x_{0},x_{2n},x_{2n-1})
\end{equation}
where $d$ is the number of dimensions, $T_{2n,2m-1}$ is the matrix used to relate modes $x_{n}$ of the string with half string modes, $\bar{x}=x(\pi/2)$ is the string midpont that can be written in terms of $x_{0},x_{2n}$ as
\begin{equation}
	x_{0}=\bar{x}+\sqrt{2}\sum_{n=1}^{\infty}x_{2n}(-1)^{n+1}
\end{equation}
and finally $\vartheta$ is a parameter having the dimension of an area.\\
Using this set of coordinates $(\bar{x},x_{2n},p_{2n})$, Witten's product among two string fields becomes the usual Moyal product as shown for the first time in \cite{[33]}:
\begin{equation}	(\Psi*\Phi)(\bar{x},x_{2n},p_{2n})=\Psi(\bar{x},x_{2n},p_{2n})\exp\Big[i\frac{\vartheta}{2}\sum^{\infty}_{n=1}\delta^{\mu\nu}\Big(\frac{\overleftarrow{\partial}}{\partial x^{\mu}_{2n}}\frac{\overrightarrow{\partial}}{\partial p^{\nu}_{2n}}-\frac{\overleftarrow{\partial}}{\partial p^{\nu}_{2n}}\frac{{\overrightarrow{\partial}}}{{\partial x^{\mu}_{2n}}}\Big)\Big]\Phi(\bar{x},x_{2n},p_{2n})
\end{equation}
In particular the noncommutativity is independent for each mode and we obtain this result from the insertion of the matrix $T_{2n,2m-1}$ in the Fourier transform. So we have
\begin{equation}
	[x^{\mu}_{2n},p^{\nu}_{2m}]_{*}=i\vartheta\delta^{\mu\nu}\delta_{nm} ~;~~ [x^{\mu}_{2n},x^{\nu}_{2m}]_{*}=0~;~~ [p^{\mu}_{2n},p^{\nu}_{2m}]_{*}=0
\end{equation}
In particular if we consider $2N$ modes, we have a noncommutative space of dimension $2Nd$.\\
\\
This version of VSFT is called Moyal String Field Theory (MSFT).
\subsection{Monoid algebra}
\label{sec:MonoidAlgebra}
We are of course interested in having the Moyal form of the sliver and wedge states in order to apply our construction. In particular the sliver state is obtained starting from wedge states in the Moyal form and performing the same operations done in Fock formalism, as we recall later.\\
First of all we analyse in details a subalgebra of $\A$, containing only Gaussian functions. It has the good property of being a monoid, and moreover it turns out to be very useful to determine the Moyal form of string fields like wedge states and the sliver which has indeed a Gaussian form. We will review the basic construction without the cutoff, and only at the end we will see how our results change when we take only a finite number $2N$ of coordinates.\\
If we consider string fields having the form of a shifted gaussian
\[\Psi_{\nn,M,\lambda}=\nn\exp(-\xi_{i} M_{ij}\xi_{j}-\xi\lambda)\]
where $\xi^{\mu}_{i}=(x^{\mu}_{2},x^{\mu}_{4},\ldots,p^{\mu}_{2},p^{\mu}_{4},\ldots)$, $M_{ij}$ is an infinite dimensional symmetric constant matrix, $\lambda^{\mu}_{i}$ is an infinite dimensional vector and $\nn$ is a normalization factor, then we have the following multiplication rules:
\begin{equation}
	\nn_{1}e^{-\xi M_{1}\xi-\lambda_{1}\xi}*\nn_{2}e^{-\xi M_{2}\xi-\lambda_{2}\xi}=\nn_{12}e^{-\xi M_{12}\xi-\xi\lambda_{12}}
\end{equation}
so it is clear that they form a closed algebra under Moyal product:
\begin{equation}
	\Psi_{\nn_{1},M_{1},\lambda_{1}}*\Psi_{\nn_{2},M_{2},\lambda_{2}}=\Psi_{\nn_{12},M_{12},\lambda_{12}}
\end{equation}
If we define
\begin{equation}
	m_{1}=M_{1}\sigma~;~~m_{2}=M_{2}\sigma~;~~m_{12}=M_{12}\sigma~~
\end{equation}
with
\begin{displaymath}
\sigma=
\left( \begin{array}{cc}
0 & i\vartheta \\
-i\vartheta & 0\\
\end{array} \right)
\end{displaymath}
then we get:
\begin{equation}
m_{12}=(m_{1}+m_{1}m_{2})(1+m_{1}m_{2})^{-1}+(m_{2}-m_{1}m_{2})(1+m_{1}m_{2})^{-1}
\end{equation}
\begin{equation}
\lambda_{12}=(1-m_{1})(1+m_{1}m_{2})^{-1}\lambda_{2}+(1+m_{2})(1+m_{1}m_{2})^{-1}\lambda_{1}
\end{equation}
\begin{equation}
\nn_{12}=\frac{\nn_{1}\nn_{2}}{\det(1+m_{2}m_{1})^{d/2}}e^{\frac{1}{4}((\lambda_{1}+\lambda_{2})\sigma(m_{1}+m_{2})^{-1}(\lambda_{1}+\lambda_{2})-\lambda_{12}\sigma (m_{12})^{-1}\lambda_{12})}
\end{equation}
It is easy to see for example that the state $\Psi_{1,0,0}=1$ corresponds to the identity of the algebra.\\
Many elements of this algebra have an inverse under the star product, but not all of them, that's why we have only a monoid structure rather than a group structure.
\subsection{Construction of wedge and sliver states}
\label{sec:ConstructionOfWedgeAndSliverStates}
We can now use this algebra to determine the form of the sliver state in MSFT.\\
First of all we can calculate the $n-th$ *-power of a state $\Psi_{\nn,M,\lambda}$:
\begin{equation}
	(\nn e^{-\xi M\xi-\xi\lambda})^{n}_{*}=\nn^{(n)}e^{-\xi M^{(n)}\xi -\xi\lambda^{(n)}}
\end{equation}
where we have
\begin{equation}\label{eq40.ps}
	m^{(n+1)}=(m+m^{(n)}m)(1+m^{(n)}m)^{-1}+(m^{(n)}-mm^{(n)})(1+mm^{(n)})^{-1}
\end{equation}
\begin{equation}
	\lambda^{(n+1)}=(1+m^{(n)})(1+mm^{(n)})^{-1}\lambda+(1-m)(1+m^{(n)}m)^{-1}\lambda^{(n)}
\end{equation}
\begin{equation}\label{eq42.ps}	\nn^{(n+1)}=\frac{\nn^{(n)}\nn}{\det(1+mm^{(n)})^{d/2}}e^{\frac{1}{4}((\lambda+\lambda^{(n)})\sigma(m+m^{(n)})^{-1}(\lambda+\lambda^{(n)})-\lambda^{(n+1)}\sigma (m^{(n+1)})^{-1}\lambda^{(n+1)})}
\end{equation}
These formulas can be simplified as follows: if we apply a transformation that diagonalizes $m$, from eq. (\ref{eq40.ps}) we see that $m^{(n)}$ and $m^{(n+1)}$ must also be diagonal, so $m$ commutes with $m^{(n)}$. This allows us to simplify formulas in
\begin{equation}
	m^{(n+1)}=(m+m^{(n)})(1+mm^{(n)})^{-1}
\end{equation}
\begin{equation}
	\lambda^{(n+1)}=(1+mm^{(n)})^{-1}[(\lambda+\lambda^{(n)})+m^{(n)}\lambda-m\lambda^{(n)}]
\end{equation}
and the explicit solution for a generic $n$ is obtained from iteraction:
\begin{equation}\label{eq55.ps}
	m^{(n)}=\frac{(1+m)^{n}-(1-m)^{n}}{(1+m)^{n}+(1-m)^{n}}
\end{equation}
\begin{equation}
  \lambda^{(n)}=(m)^{-1}(m^{(n)})\lambda	
\end{equation}
\begin{equation}\label{eq56.ps}
	\nn^{(n)}=\frac{\nn^{n}\exp\Big[\frac{n}{4}\lambda\sigma m^{-1}\lambda-\frac{1}{4}\lambda^{(n)}\sigma(m^{(n)})^{-1}\lambda^{(n)}\Big]}{\det\Big(\frac{(1+m)^{n}+(1-m)^{n}}{2}\Big)^{d/2}}
\end{equation}
In order to get the sliver state in MSFT we can use its definition in terms of vacuum state: a generic wedge state $\vert n\rangle$ is defined in Fock formalism as
\[\vert n\rangle=(\vert 0\rangle)^{n}_{*}=\vert 0\rangle *\ldots*\vert 0\rangle\]
so if we want to have its Moyal form we have to determine $\Psi_{0}(x_{2n},x_{2n-1},\bar{x}) \leftrightarrow\vert 0\rangle$. It is given by 
\begin{equation}	\Psi_{0}(x_{2n},x_{2n-1},\bar{x})=\prod_{n=1}^{\infty}\Big(16\frac{k_{2n}}{k_{2n-1}}\Big)^{d/4}\exp\Big(-\sum_{n\geq 1}\frac{k_{2n}}{2}x_{2n}x_{2n}-\sum_{n,m\geq 1}\frac{2}{\vartheta^{2}}p_{2n}Z_{2n,2m}p_{2m}\Big)
\end{equation}
where $k_{n}$ are generic frequencies (in the case of SFT we have $k_{n}=n$) and the matrix $Z$ is given by
\begin{equation}
	Z_{2n,2m}=\sum_{k\geq 1}T_{2n,2k-1}\frac{1}{k_{2k-1}}T_{2m,2k-1}
\end{equation}
So we see immediately that the vacuum state is given by a gaussian $\Psi_{\nn_{0},M_{0},0}$ with
\begin{displaymath}
m_{0}=M_{0}\sigma=
\left( \begin{array}{cc}
0 & \frac{i\vartheta}{2}k_{e} \\
-\frac{2i}{\vartheta}Z & 0\\
\end{array} \right)
\end{displaymath}
and 
\[\nn_{0}=\Big(\frac{\det(16 k_{e})}{\det k_{o}}\Big)\]
where $k_{e},k_{o}$ are diagonal matrices whose entries are respectively the even frequencies $k_{2n}$ and the odd ones $k_{2n-1}$.\\
Now if we use formulas (\ref{eq55.ps}) and (\ref{eq56.ps}) we can determine the formula for the generic wedge state:
\begin{equation}
	W_{n}(x_{2n},p_{2n})=\frac{(\nn_{0})^{n}\exp\Big(-\xi\frac{(1+m_{0})^{n}-(1-m_{0})^{n}}{(1+m_{0})^{n}+(1-m_{0})^{n}}\sigma^{-1}\xi\Big)}{\det\Big(\frac{(1+m_{0})^{n}+(1-m_{0})^{n}}{2}\Big)^{d/2}}
\end{equation}
Now we are ready to determine the sliver. It is defined as the wedge state $W_{\infty}$, that is we have
\begin{equation}
	\Xi_{m}(x_{2n},p_{2n})=\nn_{s}e^{-\xi M_{s}\xi}\sim \lim_{n\rightarrow \infty}(\Psi_{0})^{n}=\lim_{n\rightarrow\infty}(\nn_{0}e^{-\xi M_{0}\xi})^{n}_{*}
\end{equation}
where the normalization constant is determined requiring that $\Xi_{m}$ is a projector. It is found the following result:
\begin{displaymath}
	m_{s}=M_{s}\sigma=m_{0}(m_{0}^{2})^{-1/2}=\left( \begin{array}{cc}
	0 & \frac{i\vartheta}{2}k_{e}\\
	-\frac{2i}{\vartheta}Z & 0\\
\end{array} \right)
\left(\begin{array}{cc}\sqrt{\Upsilon^{t}} & 0\\
0 & \sqrt{\Upsilon}\\
\end{array}\right)
\end{displaymath}
where 
\begin{equation}
	\Upsilon=k_{e}Tk_{o}^{-1}R
\end{equation}
and the matrices $T,R$ are those used to relate full string modes with half string modes. They satisfy a list of relations given in \cite{[35]}, in particular $TR=Id; RT=Id$. The sliver matrix $M_{s}$ is
\begin{displaymath}
 M_{s}=\left( \begin{array}{cc}
	a & 0\\
	0 & \frac{1}{a\vartheta^{2}}\\
\end{array} \right)~~;~~
m_{s}=i\left( \begin{array}{cc}
	0 & a\vartheta\\
	-\frac{1}{a\vartheta} & 0\\
\end{array} \right)
\end{displaymath}
with the matrix
\begin{equation}
  a=\frac{1}{2}k_{e}\sqrt{k_{e}Tk^{-1}_{o}R}	~~;~~~m^{2}_{s}=1
\end{equation}
So the sliver is given by
\begin{equation}
	\Xi_{m}(x_{e},p_{e})=\Big(\prod_{e>0}2^{d}\Big)\exp\Big[-\sum_{n,m}\big(x_{2n}a_{2n,2m}x_{2m}-p_{2n}\Big(\frac{1}{a\vartheta^{2}}\Big)_{2n,2m}p_{2m}\big)\Big]
\end{equation}
The normalization factor 
\[\nn_{s}=\prod_{e>0}2^{d}\]
is obtained from eq. (\ref{eq42.ps}) and from the requirement 
\[\Xi_{m}*\Xi_{m}=\Xi_{m}\]
using $m^{2}_{s}=1$.\\
\\
Another interesting state obtained setting
\begin{equation}
	a'=\frac{1}{2}k_{e}~~;~~\lambda=0~~~
\end{equation}
is the butterfly state:
\begin{displaymath}
	\left. \begin{array}{lll} \Psi_{butt}(x_{2n},p_{2n}) & = & \nn_{butt}e^{-\xi M_{butt}\xi} \\	&=&\Big(\prod_{e>0}2^{d}\Big)\exp\Big[-\sum_{n,m}\Big(x_{2n}a'_{2n,2m}x_{2m}+\frac{1}{\vartheta^{2}}p_{2n}\Big(\frac{1}{a'}\Big)_{2n,2m}p_{2m}\Big)\Big]
\end{array}\right.
\end{displaymath}
The butterfly state is another solution of VSFT equation of motion interpreted as a $D25$-brane, as discussed in \cite{[44]}, having the form of a Gaussian as seen above.\\
The construction carried over so far uses all the string degrees of freedom, that in our case are given by $\left\{x_{2n},p_{2n}\right\}$ for $n=1,\ldots,\infty$. Notice that in the solutions presented above the coordinate $\bar{x}$ does not appear. \\
The cutoff procedure consists in taking only a finite number of degrees of freedom, like $\left\{x_{2n},p_{2n}\right\}$ with $n=1,\ldots,N$ from the beginning, and then perform the same steps seen above. Of course if $N$ grows up, we get a better approximation of the original VSFT, which needs all the degrees of freedom.\\
In our work we consider a cutoff version of the theory, allowing $N$ to get any finite number. It is clear that in this case we get a noncommutative theory defined on a $2Nd$ dimensional manifold, where $d=26$.\\
Now let us consider a cutoff $2N$ for each dimension. In this case we have $2N\times 2N$ dimensional matrices $M_{s}^{(2N)},M_{butt}^{(2N)}$, also the normalization factor becomes finite:
\begin{equation}
	\nn_{s}=\nn_{butt}=\prod_{e>0}^{2N}2^{d}
\end{equation}
so we get
\begin{equation}
	\Xi^{(2N)}_{m}(x_{2n},p_{2n})=\Big(\prod_{e>0}^{2N}2^{d}\Big)\exp\Big[-\sum_{n,m>0}\big(x_{2n}a_{2n,2m}^{(N)}x_{2m}-p_{2n}\Big(\frac{1}{a\vartheta^{2}}\Big)_{2n,2m}^{(N)}p_{2m}\big)\Big]
\end{equation}
\begin{equation}
	\Psi^{(2N)}_{butt}(x_{2n},p_{2n})=\Big(\prod_{e>0}^{2N}2^{d}\Big)\exp\Big[-\sum_{n,m>0}\Big(x_{2n}{a'}_{2n,2m}^{(N)}x_{2m}+\frac{1}{\vartheta^{2}}p_{2n}\Big(\frac{1}{a'}\Big)_{2n,2m}^{(N)}p_{2m}\Big)\Big]
\end{equation}
In particular matrices $a_{2n,2m}^{(N)},{a'}_{2n,2m}^{(N)}$ used both in the sliver and the butterfly case have positive definite eigenvalues: it is easy to see it in the butterfly case, while for the sliver we can show both numerically and analitically that the inverse $\Gamma$ of $\Upsilon$, in the finite dimensional case, has positive eigenvalues (see \cite{[35]} p. 20). These states have finite norm.\\
\\
This result is very encouraging, as both the cutoff sliver and butterfly have the form of gaussian functions over a noncommutative $\RR^{2Nd}$ manifold and their product is given by the usual Moyal product
\footnote{The infinite dimensional case $2N\rightarrow\infty$ is more involved and it has to be dealt with much more care: for example the matrix $\Upsilon$, after diagonalization has all positive definite eigenvalues, while its transpose $\Upsilon^{t}$ has a zero eigenvalue, while all the others are shared with $\Upsilon$, so we get a "flat" direction in the gaussian.\\
Moreover the spectrum of matrices $a_{2n,2m}$ is continuous, so it must be understood what is the meaning of a gaussian defined by means of an infnite dimensional matrix $M$ with continuous spectrum. Moreover the normalization factor becomes infinite, so the normalization also must be understood. We leave these problems for future work.}.
We know that Gaussians belong to the space $\Sc(\RR^{2Nd})$ of Schwartzian functions, which is a pre $\Cs$-algebra under Moyal's product (see \cite{[38],[39]}), so we can consider its norm completion $\overline{\Sc}(\RR^{2Nd})$ and study the K-theory group associated to such $\Cs$-algebra. This gives a mathematical classification in the cutoff theory of solutions seen.

\section{Some mathematical tools}
\label{sec:MathematicalPreliminaries}
In this section we recall some of the mathematical tools involved in our construction. In particular we review the relation between the $\Cs$-algebra given by the completion of Schwartzian functions and the $\Cs$-algebra of compact operators.
\subsection{The relation between $\overline{\Sc}(\RR^{2Nd})$ and compact operators}\label{sec2.2}
\label{sec:TheSpaceScRR2nAndCompactOperators}
When we have a generic $\Cs$-algebra, given for example by a space of functions or operators, and we want to determine its K-theory group, it can be a good strategy to see if such algebra is isomorphic to some $\Cs$-algebra of operators, as K-theory groups of many operator algebras are known, otherwise it could be very hard to find them.\\
\\
In our case we have that
\[\overline{\Sc}(\RR^{2Nd})\simeq \Bc\Big(\bigotimes^{Nd}_{a=1}L^{2}(\RR)\Big)\simeq \Bc(L^{2}(\RR^{Nd}))\]
the $\Cs$-algebra of compact operators, where we used the isomorphism $L^{2}(\RR^{n})\simeq \bigotimes^{n}_{i=1}L^{2}(\RR)$.\\
To obtain this we have to use the famous relation between operators acting on a Hilbert space and functions, given by the Weyl-Wiegner construction (see \cite{[3],[4]}).\\
The idea is the following: bounded operators $\Bo(\HH)$ on a Hilbert space have a composition law which is associative but non commutative: in general
\[\mathcal{O}_{1}\circ\mathcal{O}_{2}\neq\mathcal{O}_{2}\circ\mathcal{O}_{1}\]
We would like to have the same property for functions defined over a commutative space. This brings to the definition of Moyal's product. Infact we consider a functional space $\mathcal{F}$ and define a map 
\[\Omega: \mathcal{F}\longrightarrow \mathfrak{B}(\mathcal{H})\]
where $\mathcal{H}$ is a Hilbert space suitably defined.\\
There are various possible definitions of the map $\Omega$, each focusing on different characteristics. In particular we are interested in Weyl's map $\Omega_{W}$ which is a correspondence that associates to every function $f:\RR^{2n}\rightarrow\C$ an operator-valued function of self-adjoint operators $\hat{x}_{i}$, acting over a particular $\mathcal{H}$ and it is defined as
\[F(\hat{x})=\Omega_{W}(f):=\int d^{2n}x~ f(x)\int d^{2n}\xi~e^{-i\xi\centerdot(\hat{x}-x)}\]
This can also be written as
\begin{equation}
	F(\hat{x})=\Omega_{W}(f)(\widehat{p},\widehat{q})=\int d^{n}\sigma~ d^{n}\tau~ d^{n}p~ d^{n}q~ e^{-i\tau(\widehat{q}-q)-i\sigma(\widehat{p}-p)}f(p,q)
\end{equation}
where we have set
\[x=(q,p)~~;~~\xi=(\sigma,\tau)~~\]
There exists an inverse too, known as the Wiegner map defined by
\[f(x)=\Omega^{-1}_{W}(F)=\pi^{n}\mathrm{Pfaff}(\vartheta)\int \frac{d^{2n}\xi}{(2\pi)^{2n}}~\mathrm{Tr}_{\mathcal{H}}\Big(F(\hat{x})~ e^{-i\xi\centerdot(\hat{x}-x)}\Big)\]
Now we have that the map $\Omega_{W}$ has the following property:
\begin{equation}\label{eq69.ps}
	\Omega_{W}(f^{*})=(\Omega_{W}(f))^{*}
\end{equation}
as shown in the Appendix.\\
The same holds for the inverse map $\Omega^{-1}_{W}(f)$.\\
\\
The two maps are defined so to have the following properties:
\[\Omega_{W}(f\star g)=\Omega_{W}(f)\circ \Omega_{W}(g)~~~;~~~\Omega_{W}(f^{*})=\Omega^{*}_{W}(f)\]
\[\Omega^{-1}_{W}(F\circ G)=\Omega^{-1}_{W}(F)\star \Omega^{-1}_{W}(G)~~~;~~~(\Omega^{-1}_{W}(F))^{*}=\Omega^{-1}_{W}(F^{*})\]
where $\star$ is the Moyal product.\\
\\
Now we can apply this map to Schwartzian functions whose images belong to a subset of bounded operators:
\[ \Sc(\RR^{2Nd})\longrightarrow \mathfrak{C}\subset\Bo\]
If we consider the norm closure using the $sup$ norm on both sides we get an isomorphism of $\Cs$-algebras:
\begin{equation}
	\overline{\Sc}(\RR^{2Nd})\longrightarrow \Bc(L^{2}(\RR^{Nd}))
\end{equation}
where $\Bc$ is the set of compact operators acting on $L^{2}(\RR^{Nd})$.\\
This is a very important result as it allows us to determine the K-theory group associated to the completion of Schwartzian functions simply knowing the one associated to compact operators.
\section{K-theory from $\Cs$-algebras}
\label{sec:KTheotyFromCsAlgebras}
This part is written on behalf of readers that do not have a knowledge of K-theory groups in the framework of $\Cs$-algebra. It is far from being exhaustive, and it aims to give the taste of what is going on. A complete treatement on this subject can be found in \cite{[6]}. It can be skipped from expert readers.\\
In this section we want to analyse briefly the basic construction of K-theory groups using some special elements inside a $\Cs$-algebra, the so called \textbf{projectors}, which have some good properties as we will see now. The idea is to apply such construction to the $\Cs$-algebra of compact operators acting on a Hilbert space $\HH$, as its K-theory can give interesting physical informations.
\subsection{Projectors in $\Cs$-algebras}
\label{sec:ProjectorsInCsAlgebras}
Once we have a $\Cs$-algebra, there are some preferred elements inside it: 
\begin{definition}
A \textbf{projection} is a self-adjoint idempotent, that is
\[p=p^{*}=p^{2}\]
Two projectors $p,q$ are called \textbf{orthogonal} if $pq=0$
\end{definition}
From the above definition we easily see that the sum of two projectors is a projector precisely when they are orthogonal.\\
\\
Given all the projectors in a $\Cs$-algebra $A$ we can introduce equivalence relations. For a generic $\Cs$-algebra there are three relations among projectors and in general they do not coincide, but in the case we are considering they agree, so we will say two projectors $p,q$ to be \emph{unitarily equivalent} $p\sim_{u}q$ if $p=u^{*}qu$ being $u\in \hat{A}$ a unitary (where $\hat{A}$ is the unitization of the algebra $A$).
\begin{remark}\label{rem10}
$~~~$\\
If we consider the $\Cs$-algebra $\Bo(\HH)$ then we find that two projectors $p,q$ are equivalent iff they have the same rank, that is, if, given a Hilbert space $\HH$ on which they act, we have 
\begin{equation}
\dim p\HH=\dim q\HH	
\end{equation}
\end{remark}
Now that we have learnt how to build equivalence classes for projectors we have to learn how to add them. Then we have the following
\begin{theorem}\label{th10}
If $p_{1},p_{2},q_{1},q_{2}$ are projectors in $A$ such that $p_{1}\sim q_{1}$, $p_{2}\sim q_{2}$, $p_{1}\perp p_{2}$, $q_{1}\perp q_{2}$ then $p_{1}\oplus p_{2}\sim q_{1}\oplus q_{2}$.
\end{theorem}
In order to define properly the $K$-theory group of a $\Cs$-algebra $A$, we have to consider the pre $\Cs$-algebra given by $\Ma^{\infty}(A)=A\otimes \Ma^{\infty}(\C)=\bigcup_{n\in\N}\Ma^{n}(A)$, where $\Ma^{n}(A)$ is the set of $n\times n$ matrices with entries in $A$, given by the union of all these finite dimensional matrices. Then projectors are "embedded" in the algebra as follows:
\begin{displaymath}
p\rightarrow
\left( \begin{array}{ccc}
p & 0 & \ldots\\
0 & 0 &\ldots\\
\vdots& \vdots & \ddots
\end{array} \right)
\end{displaymath}
but if we use a unitary transformation $\left( \begin{array}{cccc}
0 & 1 & 0 & \ldots \\
1 & 0 & 0 & \ldots \\
0 & 0 & 0 & \ldots \\
\vdots & \vdots & \vdots& \ddots\\
\end{array} \right)$ then we have 
\begin{displaymath}
p\rightarrow
\left( \begin{array}{ccc}
p & 0 & \ldots\\
0 & 0 &\ldots\\
\vdots& \vdots & \ddots
\end{array} \right)\sim_{u} \left( \begin{array}{ccc}
0 & 0 & \ldots\\
0 & p &\ldots\\
\vdots& \vdots & \ddots
\end{array} \right)
\end{displaymath}
This construction is necessary to get enough orthogonal projectors to build the $K$-theory group.\\
\\
The set of all equivalence classes of projectors of a $\Cs$-algebra $A$ is an abelian monoid:
\[M(A):=\left\{[p]~s.t.~ p=p^{*}=p^{2}\in \Ma^{\infty}(A)\right\}\]
with
\[[p]+[p']:=[\textrm{diag}(p,p')]=[q\oplus q']\]
with $p\sim q$, $p'\sim q'$, $q\perp q'$ and where the identity is the class of the zero projector $[0]$.\\
\\
Now let us apply such construction to the case we are interested in: $A=\Bc(\HH)$ where $\HH$ is a separable Hilbert space (in our case we have $\HH=L^{2}(\RR^{Nd})$). In this case we use the fact that $\Ma^{\infty}(\Bc(\HH))\simeq \Bc(\HH)$. Then from remark (\ref{rem10}) we have that projectors are equivalent if they have the same rank. In particular every projector in $\Bc(\HH)$ is a finite rank projector (this is not true for $\Bo(\HH)$ as the identity is a projector in such algebra with infinite rank), so we have
\[M(\Bc(\HH))=\N\cup\left\{0\right\}\]
\subsection{The $\K(\Bc(\HH))$ group}
\label{sec:TheKAGroup}
Starting from a monoid we can build the enveloping Grothendieck group for $M(A)$; it is a commutative group $K_{0}(A)$ whose elements are given by the formal differences of elements in $M(A)$: given $[p],[q]\in M(A)$, then $[p]-[q]\in K_{0}(A)$.\\
In particular we can map elements of $M(A)$ in $K_{0}(A)$ as
\[i_{A}([p]):=[p]-[0]\]
Actually we have to pay attention, infact we can apply this construction both for unital and non-unital $\Cs$-algebras, but the naive construction just outlined works only for the unital case, while for non-unital we need a more general construction, involving the unitization $\hat{A}$ of the $\Cs$-algebra $A$. Unfortunately the $\Cs$-algebra $\Bc(\HH)$ we are interested in is not unital, so we need a slightly more difficult construction. We will not enter into the details which are quite technical. They can be found in \cite{[6]}.\\
We will simply quote the final result which is the following:
\[K_{0}(\Bc(\HH))=\Z\]
and it is the result we are interested in.
\section{Application of K-theory to VSFT}
\label{sec:ApplicationOfKTheoryToVSFT}
In this last section we want to apply the above discussion to the classification of $D25$-branes configurations. The idea is to start from $Dp$-branes considered as objects in usual string theory, to give a classification of various configurations of branes that can be physically meaningful (especially when we consider superstring theories rather than bosonic theories), and finally to check if we find the same classification when considering $D25$-branes configurations in VSFT. So let us start by defining a classification for $Dp$-branes in usual string theory.\\
\\
$Dp$-branes are very peculiar objects: they can be seen as hypersurfaces in spacetime carrying gauge theories over their worldvolume $\Sigma$. In particular if we pile up $N$ $Dp$-branes we obtain a $U(N)$ gauge theory over a $(p+1)$ dimensional manifold, so that we can define a corresponding vector bundle over their worldvolume (see \cite{[51]}). As a consequence it is possible to characterize a system of $N$ $Dp$-branes using the $U(N)$ vector bundle $E$ over the worldvolume $\Sigma$. Classes of isomorphic vector bundles over a surface can be classified using topological K-theory, which is a sort of cohomological theory for vector bundles, and which seems to be more appropriate than usual cohomology or homology theory (for a review on topological K-theory see \cite{[12]} for example). So we decide to classify $Dp$-branes configurations using topological K-theory groups. \\
$D25$-branes are a particular case where the hypersurface $\Sigma$ fills up all spacetime.\\
If we focus on bosonic string theory, as anticipated in the introduction, this classification can be considered as a \emph{mathematical} classification of $Dp$-branes as it is difficult to give a physical meaning to the use of K-theory: bosonic $Dp$-branes are unstable, as they carry a tachyonic field over their worldvolume, and so all of them are supposed to decay into the closed string vacuum. In this framework it seems hard to find room for non equivalent configurations. \\
Mabye, as suggested in \cite{[41]} there exist configurations of unstable $Dp$-branes which are not equivalent due to topological reasons and this could give a physical meaning to the use of K-groups even in a bosonic context.\\
But if we turn to superstring theories it is well known (\cite{[11],[12]}) that this construction is physically justified: $Dp$-branes are charged under RR-fields and the elements of K-theory group turn out to be the right tool to classify the RR-charges.\\
\\
As K-theory feels the topology of the branes and the branes feel the topology of spacetime, and depending on the topology the K-theory can be very complicated, we focus on a particularly simple case that is when we have a flat spacetime with flat $Dp$-branes. Infact it is a well known result that in the case of a contractible space like a flat space, all vector bundles which have the same rank are isomorphic, so $K(\Sigma)=\Z$. This means that systems of flat $Dp$-branes are classified by the number of branes that are piled up. Notice that in this classification we can also consider virtual differences of $Dp$-branes, although it is not clear the physical meaning of this operation.\\
Anyway, this is the classification of bosonic $Dp$-branes that we encounter if we consider them as surfaces carrying vector bundles, and this description is obtained in the usual string theory.\\
As already pointed out the aim of this work is to verify if we obtain the same classification (which in the case of flat spacetime is particularly simple) for systems of $D$25-branes solutions of VSFT. In particular we have found that if we take a cutoff version of VSFT described with MSFT, squeezed states solutions like the sliver and the butterfly (which in the full VSFT are interpreted as single $D25$-branes configurations) are described by Gaussian functions over a flat manifold $\RR^{2Nd}$, and their interaction is given by Moyal's product.
So their expressions are given by rank one projectors that can be classified by means of K-theory group $K_{0}(\Bc)=\Z$, and it agrees with the one found in first quantized string theory: $K_{0}(\Bc)=K(\Sigma)=\Z$, with $\Sigma$ the hypersurface representing the $D25$-brane.\\
This is an important result as objects found in VSFT, or better their cutoff version, are classified in the same way as $D25$-branes in usual string theory, and this can be another confirmation that projectors considered can be interpreted as $D25$-branes. Notice that the two classifying K-theory groups arise in completely different ways, one based on vector bundles, the other one based on $\Cs$-algebras.\\
\\
Results found in this work can be generalized in different directions:\\
\\
$\bullet$ as already said, this wants to be a preliminary study in the context of bosonic string field theory, although the real interest is to try to apply a similar construction to the case of Super String Field Theory. In particular we want to check if we manage to find solutions in the super string version of VSFT whose classification in terms of algebraic K-theory groups agree with the one usually implemented for $Dp$-branes in superstring theories, based on topological K-theory;\\
\\
$\bullet$ moreover it is interesting to check what happens when we remove the cutoff considering all the infinite degrees of freedom of the string: we expect to find the same classification found here;\\
\\
$\bullet$ in this work we considered only the case of $D25$-branes, but we know that VSFT admits solutions interpreted as lower dimensional $Dp$-branes. It would be interesting to check if we can classify them too in terms of K-theory.
Actually, using slightly different techniques it is possible to say something also in the case of lower dimensional $Dp$-branes, this is a work in progress;\\
\\
$\bullet$ finally we want to emphasise that the result found here holds in the special case of flat spacetime. It would be interesting to check if we manage to find the same agreement between $Dp$-branes for string theories and solutions of VSFT defined on non trivial spacetimes. The simplest case we should consider is the torus.\\
\\
\large\textbf{Acknowledgments}\\
\\
\normalsize The author is grateful to F. Lizzi and U. Bruzzo for the very helpful discussions and correspondence. A particular acknowledgment to professor L. Bonora and G. Landi for the reading of the work and very useful comments and suggestions. 
%
%

%

\begin{thebibliography}{99}
\bibitem{[2]} F. Lizzi, R.J. Szabo, A. Zampini: \emph{"Geometry of the gauge algebra in noncommutative Yang-Mills theory"}, \textbf{JHEP 0108} (2001) 032, [\textbf{hep-th/0107115}];
\bibitem{[3]} H. Weyl: \emph{"The theory of groups and quantum mechanics"}, Dover, 1931;
\bibitem{[4]} E.P. Wiegner, \textbf{Phys. Rev. 40}, 1932, p. 749;
\bibitem{[5]} N.P. Landsman: \emph{"Lecture notes on $\Cs$-algebras, Hilbert $\Cs$-modules, and quantum mechanics"}, \textbf{math-ph/9807030};
\bibitem{[6]} N.E. Wegge-Olsen: \emph{"$K-$theory and $\Cs$-algebras"}, Oxford University Press, 1993;
\bibitem{[7]} J.C. Pool, \textbf{J. Math. Phys. 7}, 1966, p. 66;
\bibitem{[8]} G.S. Agarwal, E. Wolf, \textbf{Phys. Rev. D2}, 1970, p. 2161;
\bibitem{[9]} J.M. Garcia-Bondia, J.C. Varilly, H. Figueroa: \emph{"Elements of noncommutative geometry"}, Birkhauser, Boston, 2000;
\bibitem{[10]} Follan: \emph{"Harmonic analysis in phase space"}, Princeton University Press, Princeton, New Jersey, 1989;
\bibitem{[1]} A. Parodi: \emph{"Toward the construction of a $\Cs$algebra in string field theory"}, \textbf{hep-th/0302177};
\bibitem{[11]} E. Witten: \emph{"D-branes and K-theory"}, \textbf{JHEP 9812} (1998) 019, [\textbf{hep-th/9810188}];
\bibitem{[12]} K. Olsen, R. J. Szabo: \emph{"Constructing $D$-branes from K-theory"}, \textbf{Adv. Theor. Math. Phys. 3} (1999) 889-1025, [\textbf{hep-th/9907140}];
\bibitem{[13]} K. Ohmori: \emph{"A review on tachyon condensation in open string field theory"}, \textbf{hep-th/0102085};
\bibitem{[14]} L. Rastelli, A. Sen, B. Zwiebach: \emph{"Classical solutions in string field theory around the tachyon vacuum"}, \textbf{Adv. Theor. Math. Phys. 5} (2002) 393-428, [\textbf{hep-th 0102112}];
\bibitem{[15]} L. Rastelli, A. Sen, B. Zwiebach: \emph{"Half string, projectors and multiple D-branes in VSFT"}, \textbf{JHEP 0111} (2001) 035, [\textbf{hep-th 0105058}];
\bibitem{[17]} A. Sen: \emph{"Universality of the tachyon potential"}, \textbf{JHEP 9912} (1999) 027, [\textbf{hep-th 9911116}];
\bibitem{[18]} A. Sen: \emph{"Descent relations among bosonic $D$-branes"}, \textbf{Int. Jour. Mod. Phys. A14} (1999) 4061, [\textbf{hep-th 9902105}];
\bibitem{[19]} L. Rastelli, A. Sen, B. Zwiebach: \emph{"String field theory around the tachyon vacuum"}, \textbf{Adv. Theor. Math. Phys. 5} (2002) 353-392, [\textbf{hep-th 0012251}];
\bibitem{[20]} E. Witten: \emph{"Non commutative geometry and string field theory"}, \textbf{Nucl. Phys. B268} (1986) p.253;
\bibitem{[21]} V.A. Kostelecky, S. Samuel: \emph{"On a nonperturbative vacuum for the open bosonic string"}, \textbf{Nucl. Phys. B336} (1990) 263;
\bibitem{[22]} A. Sen, B. Zwiebach: \emph{"Tachyon condensation in string field theory"}, \textbf{hep-th 9912249};
\bibitem{[23]} W. Taylor: \emph{"$D$-brane effective field theory from string field theory"}, \textbf{Nucl. Phys. B585} (2000) 171-192, [\textbf{hep-th 0001201}];
\bibitem{[24]} N. Moeller, W. Taylor: \emph{"Level truncation and the tachyon in open bosonic string field theory"}, \textbf{hep-th 0002237};
\bibitem{[25]} R. de Mello, A. Jevicki, M. Mihailescu, R. Tatar: \emph{"Lumps and $p$-branes in open string field theory"}, \textbf{hep-th 0003031};
\bibitem{[26]} V. A. Kostelecky, R. Potting: \emph{"Analytical construction of a nonperturbative vacuum for the open bosonic string"}, \textbf{Phys. Rev. D63} (2001) 046007, [\textbf{hep-th 0008252}];
\bibitem{[27]} L. Bonora, D. Mamone, M. Salizzoni: \emph{"$B$ field and squeezed states in Vacuum String Field Theory"}, \textbf{Nucl. Phys. B630} (2002) 163-177, [\textbf{hep-th 0201060}];
\bibitem{[28]} L. Bonora, D. Mamone, M. Salizzoni: \emph{"Vacuum string field theory with $B$ field"}, \textbf{JHEP 0204} (2002) 020, [\textbf{hep-th 0203188}];
\bibitem{[29]} L. Bonora, D. Mamone, M. Salizzoni: \emph{"Vacuum string field theory ancestors of the GMS solitons"}, \textbf{JHEP 0301} (2003) 013, [\textbf{hep-th 0207044}];
\bibitem{[30]} R. Gopakumar, S. Minwalla, A. Strominger: \emph{"Noncommutative solitons"}, \textbf{JHEP 0005} (2000) 020, [\textbf{hep-th 0003160}];
\bibitem{[31]} D. Gross, W. Taylor: \emph{"Split string field theory 1"}, \textbf{JHEP 0108} (2001) 009, [\textbf{hep-th 0105059}];
\bibitem{[32]}E. Fuchs, M. Kroyter, A. Marcus: \emph{"Squeezed state projectors in String Field Theory"}, \textbf{JHEP 0209} (2002) 022, [\textbf{hep-th 0207001}];
\bibitem{[33]} I. Bars: \emph{"Map of Witten's * to Moyal's"}, \textbf{Phys. Lett. B517} (2001) 436-444, [\textbf{hep-th 0106157}];
\bibitem{[34]} I. Bars, Y. Matsuo: \emph{"Associativity anomaly in SFT"}, \textbf{Phys. Rev. D65} (2002) 126006, [\textbf{hep-th 0202030}];
\bibitem{[35]} I. Bars, Y. Matsuo: \emph{"Computing in SFT using the Moyal star product"}, \textbf{Phys. Rev. D66} (2002) 066003, [\textbf{hep-th 0204260}];
\bibitem{[36]} D. Gross, A. Jevicki: \emph{"Operator formulation of interacting string field theory 1,2"}, \textbf{Nucl. Phys. B283} (1987), 1; \textbf{Nucl. Phys. B287} (1987), 225;
\bibitem{[37]} K. Okuyama: \emph{"Ghost kinetic operator of VSFT"}, \textbf{JHEP 0201} (2002) 027, [\textbf{hep-th 0201015}];
\bibitem{[38]} J.M. Garcia-Bondia, J.C. Varilly: \emph{"Algebras of distributions suitable for phase-space quantum mechanics.1"}, \textbf{J. Math. Phys. 29} (1988), p. 869;
\bibitem{[39]} V. Gayral, J.M. Garcia-Bondia, B. Iochum, T. Schucker, J.C. Varilly: \emph{"Moyal planes are spectral triples"}, \textbf{hep-th 0307241};
\bibitem{[40]} M. Schnabl: \emph{"String Field Theory at large $B$ field and non commutative geometry"}, \textbf{JHEP 0011} (2000) 031, [\textbf{hep-th 0010034}];
\bibitem{[41]} J. Harvey, G. moore: \emph{"Non commutative tachyons and K-theory"}, \textbf{J. Math. Phys. 42} (2001) 2765-2780, [\textbf{hep-th 0009030}];
\bibitem{[42]} E. Witten: \emph{"Non commutative tachyons and String Field Theory"}, [\textbf{hep-th 0006071}];
\bibitem{[44]} D. Gaiotto, L. Rastelli, A. Sen, B. Zwiebach: \emph{"Ghost structure and closed strings in string field theory"}, \textbf{Adv. Theor. Math. Phys. 6} (2003) 403-456, [\textbf{hep-th 0111129}];
\bibitem{[45]} D. Gaiotto, L. Rastelli, A. Sen, B. Zwiebach: \emph{"Star algebra projectors"}, \textbf{JHEP 0204} (2002) 060, [\textbf{hep-th 0202151}];
\bibitem{[46]} M. Schnabl: \emph{"Anomalous reparametrization and butterfly states in string field theory"}, \textbf{Nucl. Phys. B649} (2003) 101-129, [\textbf{hep-th 0307241}];
\bibitem{[47]} K. Okuyama: \emph{"Ratio of tensions from vacuum string field theory"}, \textbf{JHEP 050} (2002) 0203 , [\textbf{hep-th 0307241}];
\bibitem{[48]} L. Rastelli, A. Sen, B. Zwiebach: \emph{"Bounduary CFT construction of $D$-branes in Vacuum String Field Theory"}, \textbf{JHEP 0111} (2001) 045, [\textbf{hep-th 0105168}];
\bibitem{[49]} H. Hata, T. Kawano: \emph{"Open string states around a classical solution in Vacuum String Field Theory"}, \textbf{JHEP 0111} (2001) 038, [\textbf{hep-th 0108150}];
\bibitem{[50]} M. Kontsevich: \emph{"Deformation quantization of Poisson manifolds"}, \textbf{q-alg 9709040};
\bibitem{[51]} W. Taylor: \emph{"Lectures on $D$-branes, gauge theory and M(atrices)"}, \textbf{hep-th 9801182};
\bibitem{[52]} R. Szabo: \emph{"$D$-branes, tachyons and K-homology"}, \textbf{Mod. Phys. Lett. A17} (2002) 2297, [\textbf{hep-th 0209210}];
\bibitem{[53]} T. Asakawa, S. Sugimoto, S. Terashima: \emph{"$D$-branes, M theory and K-homology"}, \textbf{JHEP 0203} (2002) 034, [\textbf{hep-th 0108085}];
\bibitem{[54]} T. Asakawa, S. Kobaiashi, S. Matsura : \emph{"Closed string field theory with dynamical $D$-branes"}, \textbf{JHEP 0310} (2003) 023, [\textbf{hep-th 0309074}];
\bibitem{[55]} I. Kishimoto, Y. Matsuo, E. Watanabe : \emph{"Bounduary states as exact solutions of (vacuum) closed string field theory"}, [\textbf{hep-th 0306189}];



\end{thebibliography}
\end{document}